\documentclass[aip,amssymb,reprint,superscriptaddress]{revtex4-1}

\usepackage{mathtools} 
\usepackage{graphicx}
\usepackage{epstopdf}
\usepackage{amsmath}
\usepackage{dcolumn}
\usepackage{bm}
\usepackage{bbold}
\usepackage{amssymb}
\usepackage{mathrsfs}
\usepackage{amsfonts}
\usepackage{braket}
\usepackage{tabularx}
\usepackage{setspace}
\usepackage[caption=false]{subfig}

\usepackage{float}
\usepackage{array}
\usepackage[dvipsnames]{xcolor}
\usepackage{times}
\usepackage{changepage}
\newcolumntype{P}[1]{>{\centering\arraybackslash}p{#1}}
\newcolumntype{L}[1]{>{\raggedright\let\newline\\\arraybackslash\hspace{0pt}}m{#1}}
\newcolumntype{C}[1]{>{\centering\let\newline\\\arraybackslash\hspace{0pt}}m{#1}}
\newcolumntype{R}[1]{>{\raggedleft\let\newline\\\arraybackslash\hspace{0pt}}m{#1}}

\begin{document}

	\title[]{{Casimir spring and dilution in macroscopic cavity optomechanics}} 
	
	\author{J. M. Pate}
	\email[]{Correspondence: jpate@ucmerced.edu \\ michael.tobar@uwa.edu.au}
	\affiliation{School of Natural Sciences, University of California Merced, Merced, California 95343, USA}
	
	\author{M. Goryachev}%
	\affiliation{ARC Centre of Excellence for Engineered Quantum Systems (EQUS), Department of Physics,  University of Western Australia, 35 Stirling Hwy, 6009 Crawley, Western Australia.}
	
	\author{R. Y. Chiao}%
	\affiliation{School of Natural Sciences, University of California Merced, Merced, California 95343, USA}	
	
	\author{J. E. Sharping}
	\thanks{These two authors contributed equally.}
	\affiliation{School of Natural Sciences, University of California Merced, Merced, California 95343, USA}
	
	\author{M. E. Tobar$^{\text{ b),}}$}
	\email[]{Correspondence: jpate@ucmerced.edu \\ michael.tobar@uwa.edu.au}
	\affiliation{ARC Centre of Excellence for Engineered Quantum Systems (EQUS), Department of Physics, University of Western Australia, 35 Stirling Hwy, 6009 Crawley, Western Australia.}%
	
	\date{\today}

\begin{abstract} 
\textbf{The Casimir force was predicted in 1948 as a force arising between macroscopic bodies from the zero-point energy. At finite temperatures it has been shown that a thermal Casimir force exists due to thermal rather than zero-point energy and there are a growing number of experiments that characterise the effect at a range of temperatures and distances. Additionally, in the rapidly evolving field of cavity optomechanics there is an endeavor to manipulate phonons and enhance coherence. We demonstrate a new way to achieve this through the first observation of Casimir spring and dilution in macroscopic optomechanics, by coupling a metallic SiN membrane to a photonic re-entrant cavity. The attraction of the spatially-localised Casimir spring mimics a non-contacting boundary condition giving rise to increased strain and acoustic coherence through dissipation dilution. This work invents a new way to manipulate phonons via thermal photons leading to ``in situ'' reconfigurable mechanical states, to reduce loss mechanisms and to create new types of acoustic non-linearity --- all at room temperature.}
\end{abstract}

\maketitle

The Casimir force\cite{casimir1948attraction} for macroscopic bodies arises from the unequal pressures exerted by quantum vacuum fluctuations on the inside and outside boundaries of a cavity. This occurs because there is a discrete energy spectrum within a cavity, while outside the electromagnetic energy spectrum is continuous. Thus, an inward pressure exists on the cavity because the vacuum energy density is larger outside than inside.  At finite temperatures the thermal modification to the Casimir effect increases with temperature and thus increases the range of separations between two objects where Casimir forces are observable\cite{sushkov2011observation,bressi2002measurement,klimchitskaya2009casimir, obrecht2007measurement,rodriguez2011casimir,zou2013casimir, wilson2011observation, somers2018measurement}. The Casimir force is now not just solely of academic importance, but has been recently drawing more interest in the arena of sensors, switches, amplifiers, and photonic sources\cite{stange2019building, lamoreaux2004casimir, liu2016casimir, imboden2014design, rivera2019light}. This work significantly adds to the tool kit available for such applications and opens new avenues for device manipulation as well as a new way to investigate the Casimir force.

In this work, the ability to sense the Casimir force was achieved by realising a small gap ($\sim\mu$m) between a cm-scale microwave re-entrant cavity post and 38 mm diameter SiN metal-plated membranes. A calibrated force varied the re-entrant gap to tune the microwave resonator and transition in and out of the Casimir regime, which revealed extra stiffening of the membrane (Casimir spring). Importantly, we have determined that a spatially-localised Casimir spring may act as a lossless non-contacting boundary (pinning effect), which was verified with finite element simulations of acoustic mode frequencies. Our results have shown that this ``Casimir boundary" created ``Casimir dilution" and significantly increased the acoustic $Q$-factor of the acoustic mode. Dilution has been used as an engineering technique to reduce phononic losses through the addition of strain\cite{corbitt2007optical,ghadimi2018elastic,tsaturyan2017ultracoherent}. Thus, this technique has allowed us to create a new way to engineer non-contacting boundary conditions in a lossless way using only thermal photons, where  decreasing the gap increased both the strain and $Q$-factor in the membrane. These new techniques presented in this work have high potential to create new ``\textit{in situ}'' agile programable devices, engineered to manipulate mode structures and improve resonator losses as needed at room temperature, which could have far reaching consequences. Furthermore, we also observed a new form of acoustic non-linearity near the gap spacing where the Casimir and spring constant force were equal, allowing the measurement of bistability and other enhanced non-linear effects.

\section*{Microwave cavity and Casimir spring}
A microwave re-entrant cavity is a type of 3D lumped microwave resonator, which consists of a post enclosed in a cavity, with a gap, $x$, between the end post and end-wall boundary as shown in Fig.~\ref{fig:cartoon}, and discussed in more detail in the Supplementary Materials\cite{fujisawa1958general,le2013rigorous}. The cavity resonance frequency, $\omega/2\pi$, is primarily determined by the gap and, classically, the sensitivity of the frequency shift with respect to displacement, $d\omega/dx$, is inversely proportional to the gap. The structure acts as a reliable transducer between frequency and gap size with the microwave resonance governed by a lumped $LC$ circuit model $\omega=\left [\sqrt{L C}\right ]^{-1}$\cite{fujisawa1958general,tobar2000accurate} where the re-entrant cone acts as an inductor, $L$, and the gap behaves as a capacitor, $C$.  Thus, if the end wall is constructed as a mechanical oscillator, the system may be configured as a sensitive optomechanical device\cite{aspelmeyer2014cavity}, increasing in sensitivity as the gap becomes smaller. For example, if the end wall vibrates or is displaced, a frequency shift of the microwave mode is produced. From this effect very sensitive devices have been realised previously for a range of applications, such as the readout for a resonant-mass gravitational wave detector\cite{Blair:1995zr,tobar2000accurate}, transducers to read out induced displacement at high sensitivity\cite{Barroso:2004aa,Carvalho14,Goryachev:2015aa}, and the realisation of tuneable microwave cavities for a variety of applications\cite{Carvalho16,Davis18,pate2018electrostatic}.

Our system was constructed with a membrane as the end wall and by applying an electrostatic force we achieved exquisite positioning of the re-entrant gap \cite{pate2018electrostatic}, The gap size was numerically obtained as a function of the applied electrostatic force using the LC model. When the re-entrant gap was sufficiently large so it did not experience the Casimir force, we observed normal behavior as expected for a macroscopic cavity optomechanical system (denoted as the ``free" state) and the acoustic membrane had an unperturbed static spring constant of $k_S$. For small gaps, inside the Casimir region the Casimir force became large compared to the membrane restoring force, causing reduced motion directly under the re-entrant post with an effective spring constant of $k_C(x)$  (Casimir spring), which effectively ``pinned'' the membrane at this point due to the attractive nature of the force. This ``pinning" changed the acoustic mode shape of the membrane and hence state of the membrane (denoted as the ``pinned" state). This change in mobility is similar to the ``buckled up'' and ``buckled down'' states seen in micromechanical oscillators\cite{bagheri2011dynamic}. Interesting features were also demonstrated when the gap was spaced such that the Casimir and the spring force were of equal magnitude. Here we observed the existence of a region of increased force sensitivity for the re-entrant microwave cavity readout of the mechanical motion. These results show the most sensitive re-entrant cavities may not require the smallest gap sizes, but rather a careful balance of forces to gain the best sensitivity at room temperature. Other non-linear effects also occurred around this region, such as modifications to the driven microwave mode and optomechanical bistability indicated by the effect of mode hopping (see Supplementary Information for more experimental and device details). 

	\begin{figure}[!htb]
		\begin{center}
		\includegraphics[width=1\columnwidth]{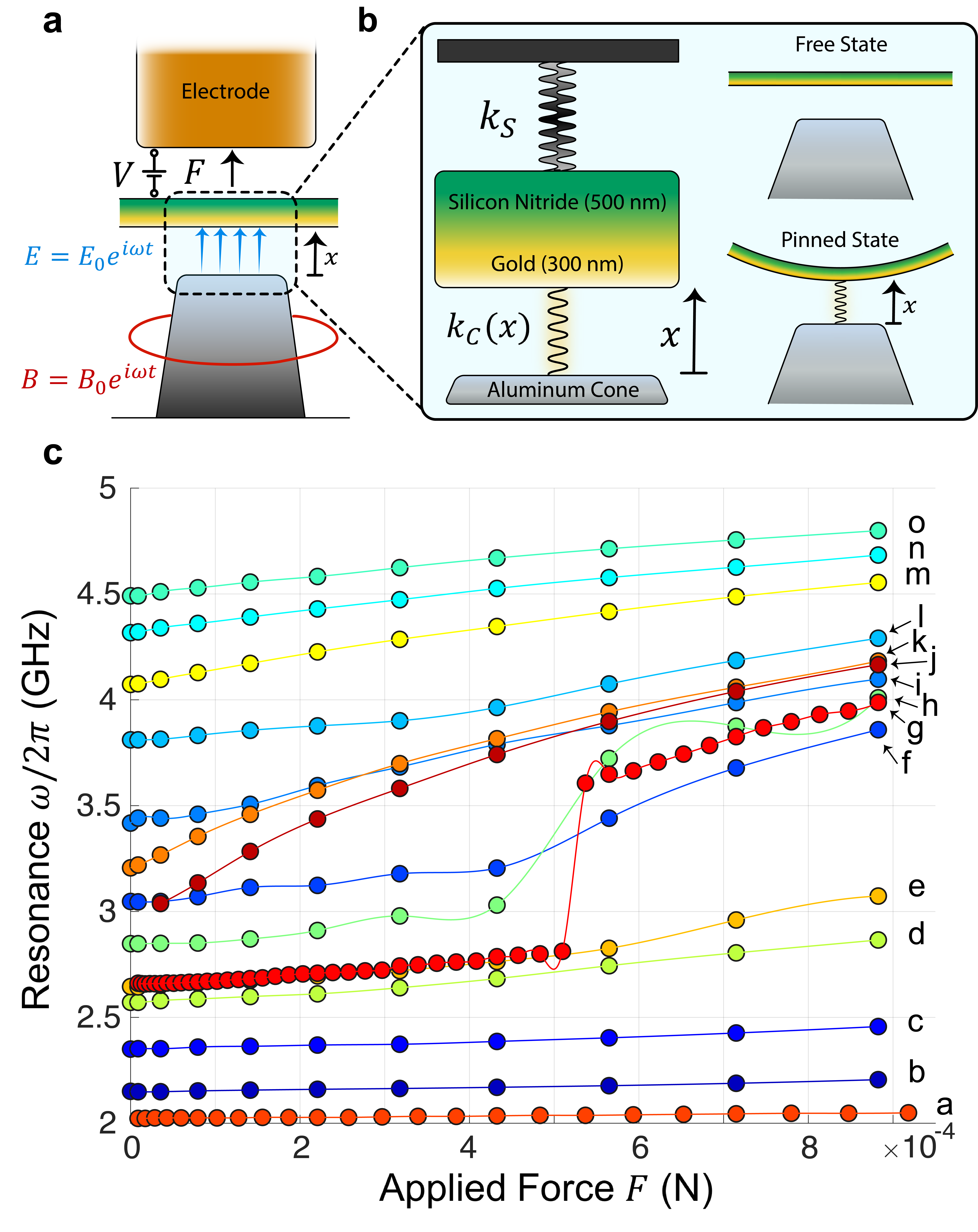}
		\caption{\textbf{a}, Diagram (not to scale) of the cross-section of the electromechanical elements of the re-entrant cavity and associated electric, $E$ and magnetic, $B$ field of the re-entrant mode. A bias voltage, $V$, on the electrode produced an attractive electrostatic force, $F$, that pulled the membrane, increasing the gap size $x$ and shifting the microwave resonance frequency, $\omega/2\pi$. \textbf{b}, The left-hand image is a simplified model of the setup illustrating the Casimir spring, $k_{C}(x)$, and the normal restoring spring, $k_{S}$, intrinsic to the membrane. The right-hand image shows the influence of the Casimir spring acting on the membrane in the vicinity of the re-entrant cone. \textbf{c}, Microwave cavity resonance frequency, $\omega/2\pi$, versus applied electrostatic force, $F$, for a range of initial gap spacings labeled $a$-$o$ (smallest gap of 0.59 $\mu m$ to largest gap of 3.9 $\mu m$). For the smallest values of initial gap spacing ($a$-$c$) the membrane experienced the thermal Casimir force significantly reducing the actuation and producing a pinned state. As the gap spacing increased ($d$-$l$) a transition to non-linear behaviour was observed. For the largest initial gap spacings ($m$-$o$) the membrane was in the free state where the Casimir response is negligible. The labels $a$-$o$ correlate with the same labels in Fig.~\ref{fig:CAS}. 
				\label{fig:cartoon}}
		\end{center}
	\end{figure}

An advantage of this macroscopic system over many other micromechanical systems used for Casimir force sensing is the capability to achieve a large dynamic range of membrane gap sizes. To understand the response of the membrane as a function of its distance from the tip of the re-entrant cone, or gap size ($x$), a thin metallic spacer was inserted on the outer edge of the cavity and a static DC voltage ranging from 0 to 250 V was applied between the external electrode and membrane. Labeled as ``$a$-$o$" in Fig.~\ref{fig:cartoon}c, the measurements were repeated for various spacer sizes in order to build a map of the cavity response. The frequency shift as a function of the electrostatic force applied by the external electrode is shown for each experimental run of voltage sweeps with splines shown for clarity. The applied force of the electrode to the membrane was calibrated by measuring the electrostatic mechanical frequency shift outside of the Casimir regime\cite{pate2018electrostatic, shoaib2016frequency}. 


The calculation of the thermal Casimir force for the given geometry is challenging due to several factors\cite{klimchitskaya2009casimir,lamoreaux2004casimir}. The 3D re-entrant cavity deviates from the typical geometric configurations of sphere-plate, sphere-sphere, or plate-plate as the re-entrant microwave mode occurs in a sub-wavelength gap size between the membrane and cone. Additionally, we note that the skin depth ($1.76 - 1.17~\mu$m for $\omega/2\pi = 2-5$ GHz) of the electromagnetic field exceeds the layer of gold roughly by a factor of five and interacts with the SiN dielectric layer. These considerations change the magnitude of the thermal Casimir force and should be considered in future investigations, but the scaling of the force as a function of gap size is preserved, and we use this scaling to unequivocally identify its nature.
%

\begin{figure}[!htb]
		\begin{center}
			\includegraphics[width=1\columnwidth]{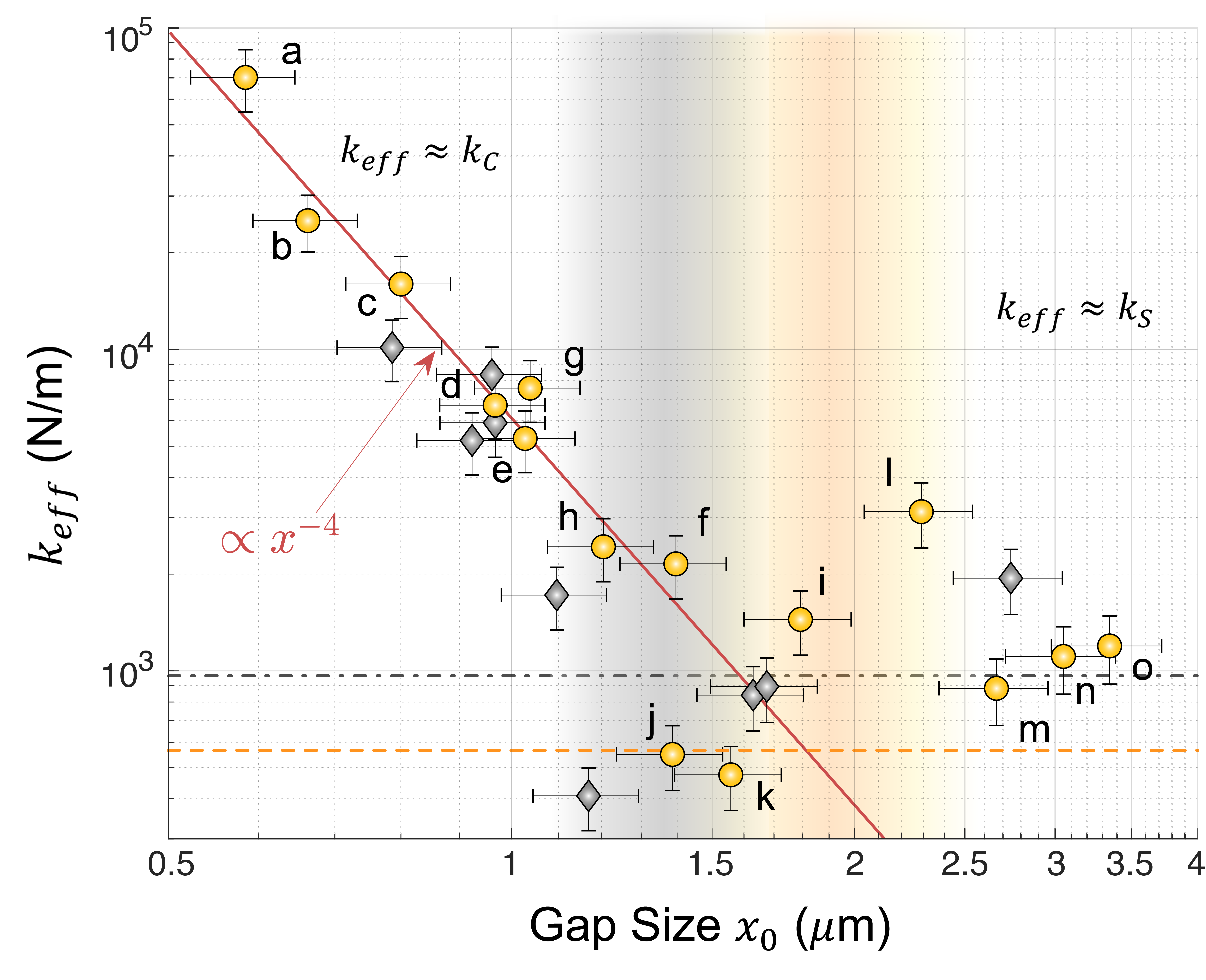}
			\caption{Thermal Casimir force graph showing effective static spring constant with inverse distance power law showing $x^{-4}$ for the gold membrane cavity (gold points) and niobium membrane cavity (gray points), corresponding to $x^{-3}$ in force. The dashed (orange) and dash-dotted (gray) flat lines represent the fundamental dynamical spring constant for the gold and niobium membranes respectively, as determined by the fundamental mode of the mechanical oscillator. The letters coincide with the lines in Fig.~\ref{fig:cartoon}c. The gold and gray shaded regions embody the non-linear zones of the mechanical resonator for the gold and niobium membranes, respectively. Error bars represent uncertainty in the cavity geometry.
				\label{fig:CAS}}
		\end{center}
\end{figure}

Figure~\ref{fig:CAS} shows the observation of the inverse-cubic power law dependence ($x^{-3}$) of the force with respect to the cavity gap size, demonstrated in the form of the effective Casimir spring. The effective static spring constant, $k_{\text{eff}}$, was obtained by considering small displacements $\delta x$ around the static value of the gap spacing, $x_{0}$, caused by the application of a small applied electrostatic force, $\delta F$, with the setup illustrated in Fig.~\ref{fig:cartoon} (see Supplementary Information for the niobium membrane data and LC circuit equations). From this measurement the effective spring constant was calculated by,
\begin{equation}
k_{\text{eff}} = \frac{\delta F}{\delta x} =  \frac{\partial F}{\partial \omega} \frac{\partial \omega}{\partial x} \Big |_{x\approx x_{0}}.
\end{equation}

Here, we used the well-determined microwave cavity frequency shift, $\delta\omega$, to calibrate the position shift of $\delta x$. Only the initial linear regime corresponding to smaller applied electrostatic forces around $x=x_0$ were considered for the calculation of $k_{\text{eff}}$. This was to prevent possible distortions of the membrane, which could occur on the application of a large force under the influence of the Casimir pinning mechanism. Any such distortions would be very minimal under a small applied force of $\delta F$.

In Fig.~\ref{fig:CAS}, the highlighted portions of the graph designate areas of non-linear behaviour and defines the gap region where the spring constant of the membrane was the same order as the Casimir spring constant. Outside this regime, for smaller gaps, spring hardening takes place ($x<$ 1.5 $\mu$m for the gold membrane and $x<$ 1.2 $\mu$m for the niobium membrane) due to the dominance of the Casimir force where the effective spring constant increases sharply as $x^{-4}$ as the gap decreases. For larger gap spacings outside the non-linear and Casimir regimes the value of the measured static spring constant was the same order as the dynamic spring constant of the fundamental frequency of the membrane. Because the niobium membrane was stiffer than the gold one, it required a larger Casimir force to enter the non-linear regime, and thus occurs at a smaller gap compared to the gold membrane.

\section*{Casimir dilution}
The dynamical response of the acoustic modes were also strongly modified due to the Casimir force. This is illustrated from measurements of the mechanical modes of the membranes using two techniques. The first was by strongly driving the acoustic modes and observing the microwave mode hopping, and the second was through the measurements of the non-driven acoustic spectrum due to Nyquist thermal fluctuations.

Mode hopping effects were observed by measuring the response of the microwave resonance while applying to the external electrode a DC bias voltage, $V_{dc}$, combined with an oscillating drive voltage, $V_{ac}(\Omega)$, of frequency $\Omega/2\pi$. The membranes were electrically driven into motion at its acoustic mode frequency, $\Omega_{m}$, when $\Omega = \Omega_{m}$. Measurements were taken for a range of gap sizes and the maximum product of the two signals $F(\Omega)\propto V_{dc}V_{ac}(\Omega)$ was held constant to preserve a steady oscillating applied force. This was a very good approximation because the external electrode gap spacing was of the order $200~\mu$m, so there was minimal change in the electrode gap over the dynamic range of the measurements. 

\begin{figure}[!htb]
		\begin{center}
			\includegraphics[width=1\columnwidth]{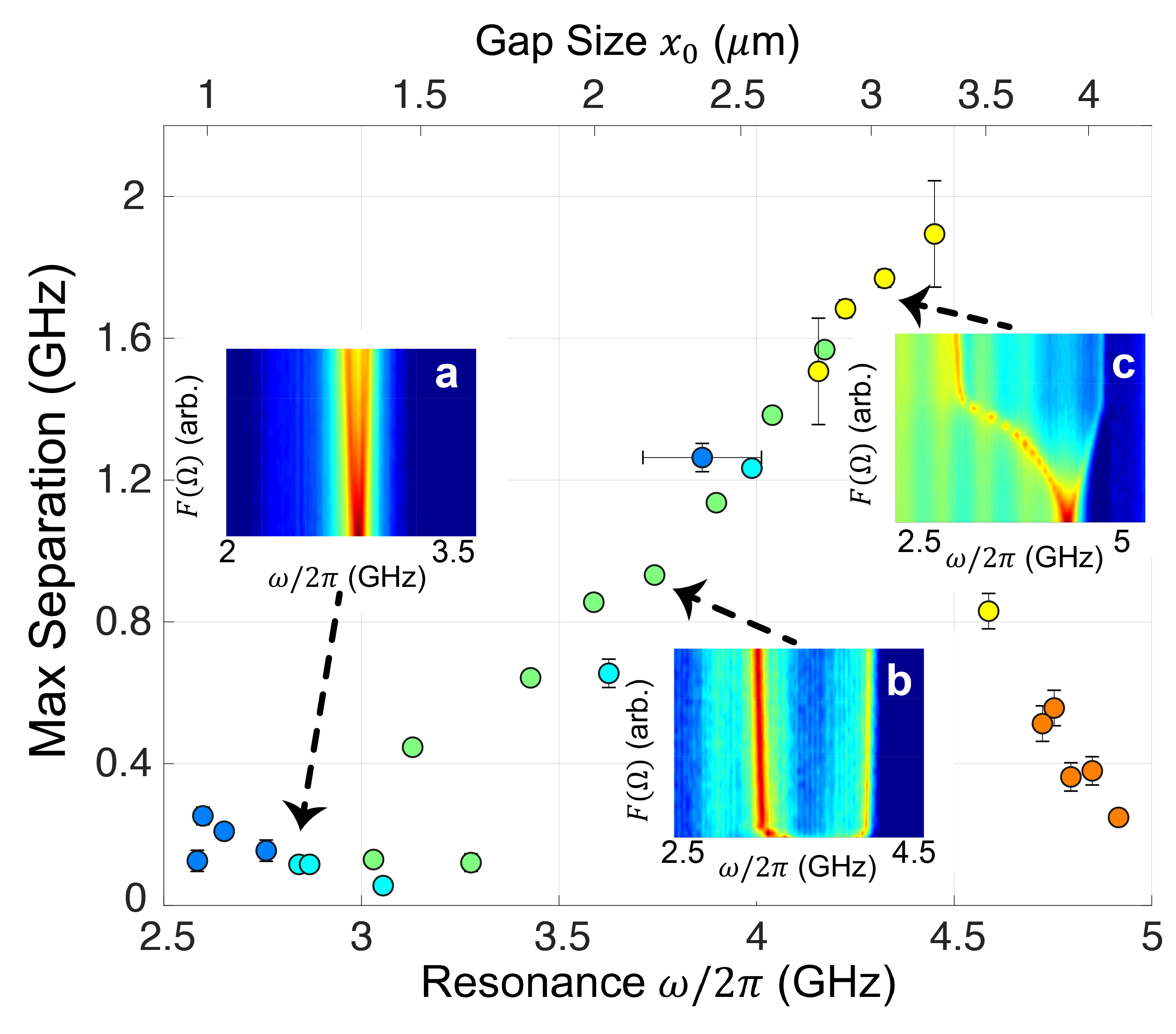}
			\caption{Five separate experimental runs of the gold membrane, showing the value of the maximum mode hopping separation of the driven microwave mode as a function of unperturbed microwave cavity frequency (lower $x$-axis label) or equivalent initial gap spacing (upper $x$-axis label). For every point the membrane experienced the same, maximum applied electrostatic force, $F(\Omega)$, and represents the maximum frequency separation between the two resonance peaks illustrated in the insets (a)-(c). The insets display the microwave transmission (dB) of the cavity as the electrostatic force was increased, leading to mode separation in three distinct regions, (\textbf{a}) the Casimir dominated region, (\textbf{b}) the non-linear region and (\textbf{c}) outside the Casimir region. Error bars represent uncertainty in the cavity geometry.
				\label{fig:MaxNMS}}
		\end{center}
	\end{figure}

Figure~\ref{fig:MaxNMS} shows the behaviour of the microwave resonance as the gold membrane was electrically driven into motion at frequencies close to $\Omega_{m}/2\pi$. As shown in the insets (a)-(c), the resonance separates in two as the applied force increases. This is because the membrane oscillates much faster than the acquisition rate of the network analyzer and appears as a separation. We describe this separation as ``mode hopping'', which is distinct from normal-mode splitting. There are three distinct regions characterized by the position of the membrane as it is electrically driven\cite{bagheri2011dynamic}. In our configuration, the region at which the cavity displays the largest mode separation does not occur for the smallest gap size due to the presence of the Casimir force. The maximum separation reaches 1.88 GHz, which is 43\% of the unperturbed cavity frequency at a gap size of $x_0=3.3 \pm 0.4$ $\mu$m. This data and the observation of mechanical bistability (see Supplementary Information) are analogous to previous results of a micromechanical oscillator\cite{chan2001nonlinear} by considering the gap size and the maximum frequency separation as a measure of oscillator amplitude.

The Casimir force tensioned the mechanical oscillator, making it stiffer so diminished microwave mode hopping was observed as shown in Fig.~\ref{fig:MaxNMS} inset (a). This is because the gold membrane is in the pinned state. Conversely, we observed the lowest threshold of mode splitting due to $F(\Omega)$ around 3.5 GHz, shown in inset (b), which is in the middle of the non-linear regime (even though the splitting is not maximum). This low threshold is a key sign that the mechanical oscillator is more susceptible to a small applied force and thus exhibits increase force sensitivity. An asymmetry is observed, shown in inset (c), which is explained by considering the region of the Casimir force. The lower frequency branch represents a smaller value of $x$, and is thus perturbed more by the influence of the Casimir force compared to the higher frequency branch.

For the niobium membrane we acquired the acoustic mode spectrum through the use of a microwave phase-bridge read-out circuit (more details given in the supplementary material). In the demodulated microwave spectrum, we distinguished the mechanical resonances from residual noise peaks by sweeping the frequency of $V_{ac}(\Omega)$ on the external electrode and observing the modes which exhibited microwave mode hopping. The multitude of mechanical resonances were not as easily characterised or observed in the company of the Casimir force due to the pinned response. However, we observe driven microwave mode hopping for acoustic frequencies of 4.85 kHz, 10.44 kHz, 16.06 kHz, 21.66 kHz, and 27.29 kHz  confirming they were membrane modes. Interestingly, these modes were not defined by the traditional modes of a clamped circular membrane. Under further investigation we verified through finite element analysis that these resonances were radially-symmetric modes, pinned in the middle of the membrane, and justified by the frequency shifts in the acoustic spectrum as they transitioned into the Casimir force region. In contrast, as the gap size moved beyond the Casimir region ($\omega/2\pi > 2.8$ GHz), into the free state, the mechanical modes were confirmed to be well-defined by the expected frequencies of a clamped, circular membrane and identified in both the microwave and acoustic spectrum (see Supplementary Information for data regarding the acoustic spectrum).

The discovery of the aforementioned effect has allowed us to utilise the Casimir force to strain engineer an acoustic membrane via dissipation dilution, enhancing the mechanical resonator $Q$-factor\cite{ghadimi2018elastic, tsaturyan2017ultracoherent}. This is illustrated in Fig.~\ref{fig:COMSOL} by the niobium membrane shifting upwards in frequency while the acoustic $Q$-factor, $Q_{m}$, increases as the gap size becomes smaller for both (0,2) and (0,3) modes. The increased tension is non-contacting, which reduces clamping losses commonly seen in mechanical resonators. Initially as the gap size became smaller and transitioned into the non-linear and Casimir regimes, the mechanical spring first softened and then hardened as the Casimir force increased, consistent with the microwave data taken for the static gap spacings. The mechanical modes also varied in quality factor between the pinned and free states. We have identified that $Q_{m}$ undergoes large swings in the non-linear shaded area. Outside of the non-linear region, $Q_{m}$ saturates to a single value for large gaps, but continues to increase under the influence of the Casimir force for small gaps. Considering the point of least stiffness there is an improvement in $Q_{m}$ by factors of 14.5 and 13.1 for the (0,2) and (0,3) modes, respectively. Moreover, the quality factor was improved by an order of magnitude beyond the bare membrane $Q_{m}$ for the (0,2) mode, and continues to rise for smaller displacements. Thus, $Q_{m}$ is expected to improve as a result of increased dilution if the gap size was further reduced. We were limited in our capability to measure smaller gaps, as in the pinned state our readout cavity is in proximity to a node of the mode shape. The addition of another resonator in the free state at the anti-node would allow improved measurements at smaller gap spacings, and the possibility of measuring a higher $Q_{m}$.

Fig.~\ref{fig:COMSOL} also shows good agreement of experimental data with the frequency shifts calculated from a COMSOL model of the membrane under the influence of the Casimir force, which is approximated as a variable-spring at the point below the re-entrant post. We chose a point spring to approximate the presence of the Casimir force because the ratio of re-entrant cone radius to cavity radius is only 1\%. The lower arrows in both panels show the points of the (0,2) and (0,3) bare mode frequencies without a spring and the upper arrows are locations at which the mode frequency appeared to saturate for an arbitrarily-large spring (see Supplementary Information). The inset images display the mode shapes for both pinned (Casimir) and free (non-Casimir) states. 

\section*{Concluding Remarks}
In summary, we have tested the narrowest gap re-entrant microwave cavities coupled to a mechanical resonator to date, and observed that the Casimir force gives rise to a plethora of interesting phenomena. For example, we have made the first observation of the Casimir spring and dilution effect in a macroscopic optomechanical system. The Casimir spring was observed by examination of the power law of the Casimir force as a function of gap size between the SiN metallic-coated membranes and a re-entrant post. This was possible because the re-entrant post formed a microwave cavity, which allowed self calibration of the gap from the readout of the microwave frequency. Casimir dilution was also witnessed from the creation of an effective non-contacting boundary condition due to the localisation of the Casimir force under the re-entrant post, pinning the movement of the membrane at that point. This caused the acoustic mode of the membrane to change from a ``free'' state to a ``pinned'' state as it transitioned into the Casimir region for small gaps. Mechanical frequency shifts occurred for all radially-symmetric modes during this transition, which also increased the mechanical quality factor by over an order of magnitude as the Casimir force was increased. The transition of the mechanical modes from the traditional circular drum model, to a mode with a pinned boundary in the middle of the drum are in excellent agreement with finite-element simulations. 

\begin{figure}[!htb]
		\begin{center}
			\includegraphics[width=1\columnwidth]{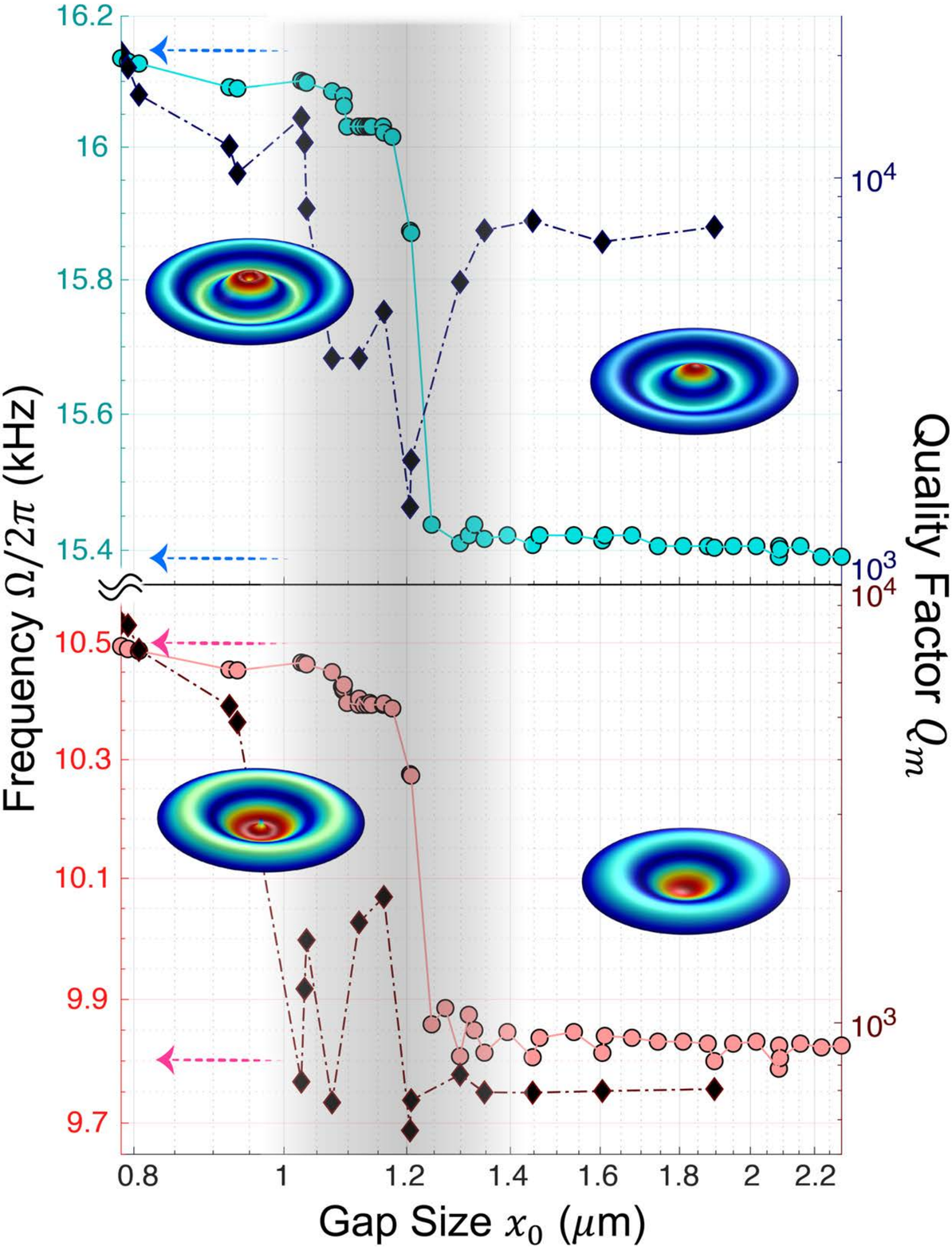}
			\caption{The effect of Casimir pinning and dilution on the (0,2) and (0,3) niobium membrane modes. The filled circles are the membrane frequencies and the filled diamonds are the corresponding mechanical quality factors. The inset images on the left show the pinning in the center of the modal structure which gives rise to the frequency shift. The inset images on the right show the mode in the free state, where the center is allowed to move. The lower arrows on both panels are the starting resonances from simulations for the (0,2) and (0,3) modes. The upper arrows for both graphs correspond to the frequency at which the simulated mode frequency saturate due to increased pinning force, in good agreement with the data. The shaded regions are areas of mechanical non-linearity, which divides the pinned state (on the left) and the free state (on the right).
				\label{fig:COMSOL}}
		\end{center}
\end{figure}

The observation of a new way to engineer non-contacting dissipation dilution and boundary conditions creates a path forward to realise unique, topological mechanical oscillators while also increasing the acoustic $Q$-factor. We anticipate this effect will have a large impact on the field of cavity optomechanics and the hybrid integration of optical, electrical, and mechanical systems\cite{bagci2014optical, ghadimi2018elastic, polzik2017optical}. For example, recently invented multiple post re-entrant cavities utilising posts and rings\cite{Goryachev:2015aa,6843369,patent2014} could be coupled to an acoustic membrane in a similar way as described here. This would allow the manipulation of many degrees-of-freedom and the coupling of multiple photon resonances for membrane manipulation and readout. These systems could also in principle be configured to strongly couple the acoustic membrane to spin systems such as magnons\cite{Goryachev:2014aa,Kostylev:2016aa} and defects in diamonds\cite{PhysRevB.91.140408}, as the multiple post re-entrant cavity has produced the strongest such couplings to date.


Furthermore, we established the existence of a very interesting non-linearity when the Casimir and membrane spring constant are of similar magnitude. Future work will further seek to explore this non-linear phenomena of the mechanical resonator and its relation to the thermal Casimir force\cite{rhoads2009nonlinear, huang2016generating}. Moreover, this work could be developed for a system, which utilises the non-linearity generated by the thermal Casimir force to enhance force sensing\cite{imboden2014design, rhoads2009nonlinear}. Additional avenues for this particular research intersect with the active field of searching for the axion particle as well as a candidate for a possible fifth fundamental force\cite{brax2007detecting, almasi2015force} because of our ability to transition in and out of the thermal Casimir regime.

\section*{References}
%

\vspace{5mm}
\vspace{5mm}

\noindent\textbf{Acknowledgments} J.M.P. thanks E. Ivanov for his help with the development and analysis of the microwave phase-bridge circuit. J.M.P. also thanks Keith Blackburn for his help with machining the vacuum testbed. This research was supported by the Australian Research Council Grant No. CE170100009 and DARPA through grant W911NF1510557. 

\noindent\textbf{Author Contributions} The experimental work was carried out by J.M.P. The manuscript and theoretical work was done by J.M.P, M.G., R.Y.C., J.E.S. and M.E.T. 

\noindent\textbf{Competing Interests} The authors declare no competing interests.

\noindent\textbf{Code availability} The data that support the plots within this paper and other findings of this study are available from the corresponding authors (JP and MET) on reasonable request.

\noindent\textbf{Data availability} Finite element analysis was undertaken using COMSOL, implemented meshes and data output may be requested from the corresponding authors (JP and MET).

\noindent\textbf{Additional Information} Supplementary Information is available in the online version of this paper. Correspondence and requests for materials should be addressed to  the corresponding authors (JP and MET).

\newpage
\onecolumngrid
\appendix
\begin{widetext}
\begin{center}
\textbf{{\Large Supplementary Methods}}
\end{center}

The real device is pictured in Fig.~\ref{fig:pic} with the three pieces that comprise the re-entrant cavity: the cavity body, the membrane, and the electrode adapter. The cavity was constructed using traditional machining methods on a lathe, CNC, and milling machine. The tolerance of the machines did not allow for the manufacturing of gap sizes of less than 10~$\mu$m. Instead, a cylinder of 6061 aluminium was cut on a horizontal band saw, faced-off with a lathe, and polished on a polishing wheel for several hours starting from fine sand paper and ending with felt. Afterwards, the aluminium cylinder was bored out on a lathe taking care never to touch the center of the polished aluminium with a tool. Final reduction of the gap to $\sim1~\mu$m was achieved by iteratively measuring the cavity resonance frequency with a network analyzer and performing light surface removal of the outer ledge using Scotch-Brite on a lathe. 

\begin{figure}[!htb]
\begin{center}
\includegraphics[width=12cm]{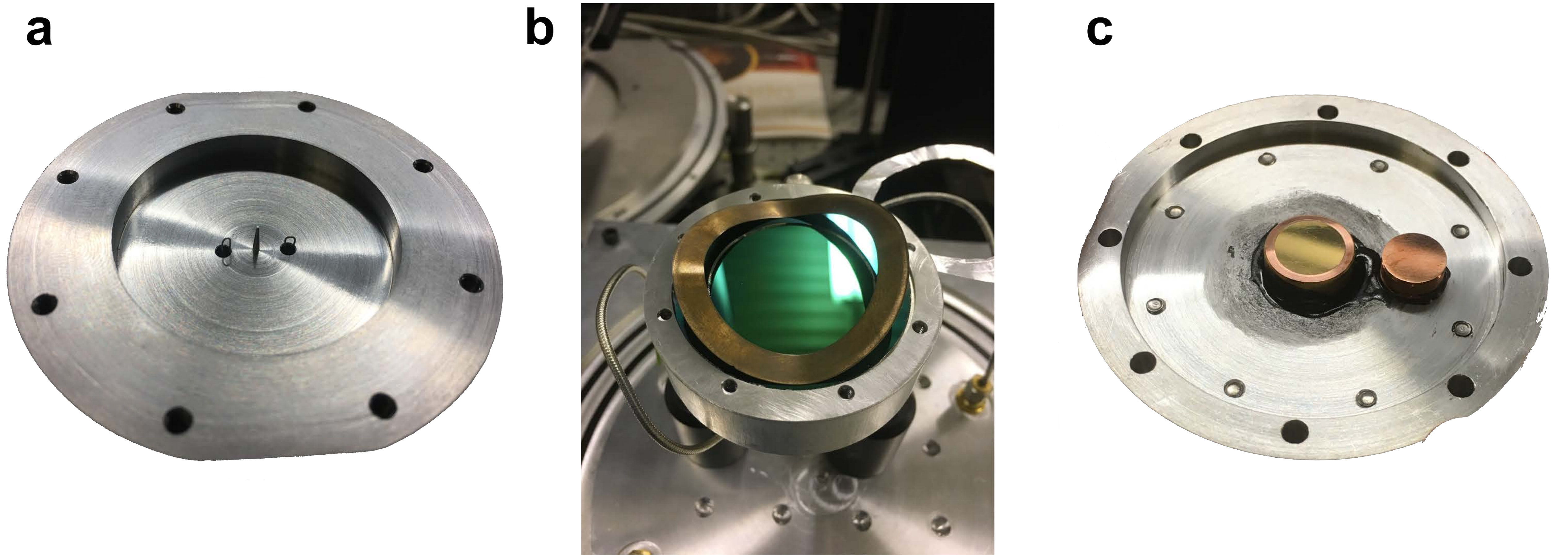}
	\caption{(\textbf{A}) The bare re-entrant cavity shown with cone and loop antennas. (\textbf{B}) The SiN membrane (gold underneath) acted as the cavity boundary and determines the resonance frequency. The bronze spring washer was pressed against the membrane and held it securely in place. A thin (1 mm) aluminum washer created the space needed to bring the electrodes close to the uncoated membrane surface. The cavity was mounted in a vacuum chamber on an optical table. (\textbf{C}) The center gold-plated electrode for actuating the membrane. The electrodes and mounting bracket were flipped over and placed above the bronze spring washer. The smaller, off-center electrode was traditionally designed to ``drive'' the membrane with the center electrode originally created to ``sense'' the vibrations, however for these experiments the second electrode was not used and the center electrode was purely used as an active device. }
	\label{fig:pic}
	\end{center}
\end{figure}

The gap size was numerically obtained as a function of the applied electrostatic force using the measurement of the re-entrant cavity frequency and the well known LC model (discussed later). Based upon the starting resonance frequency we estimated the smallest gap achieved for the gold membrane to be $x=585\pm 61$ nm. Numerical estimates yielded a frequency-pull parameter of $G/2\pi \equiv \frac{df}{dx} = 1.66 - 0.57$ GHz/$\mu$m for the range of resonances presented in this paper. The intrinsic microwave cavity quality factors ranged between $20<Q_{0}<300$ at room temperature, which is expected due to a small geometry factor of the re-entrant mode.

For the mechanical oscillator we used Silicon-Nitride (SiN) membranes (Norcada) of diameter of 38.1 mm and total thickness of 800 nm (500 nm SiN and 300 nm of metal deposited on the underside). Two different membranes were used, one with a gold and the other with a niobium coating. The fundamental frequency of the gold and niobium membranes were $\Omega_{m}/2\pi=2.51$ kHz and $\Omega_{m}/2\pi=4.3$ kHz, respectively, with effective masses of $m_{\text{eff}}= 2.3$ mg and $m_{\text{eff}}= 1.3$ mg, respectively. An external gold-coated copper electrode was placed near the uncoated surface of the membrane forming a capacitor. This electrode actuated the static mechanical motion of the membrane through a DC bias\cite{pate2018electrostatic}. The whole re-entrant cavity-membrane system was held in a vacuum chamber with microwave and DC voltage feedthroughs attached to allow control, measurement and characterization at a pressure of about $1\times10^{-3}$ mbar.

\begin{figure}[!htb]
		\begin{center}
			\includegraphics[width=12cm]{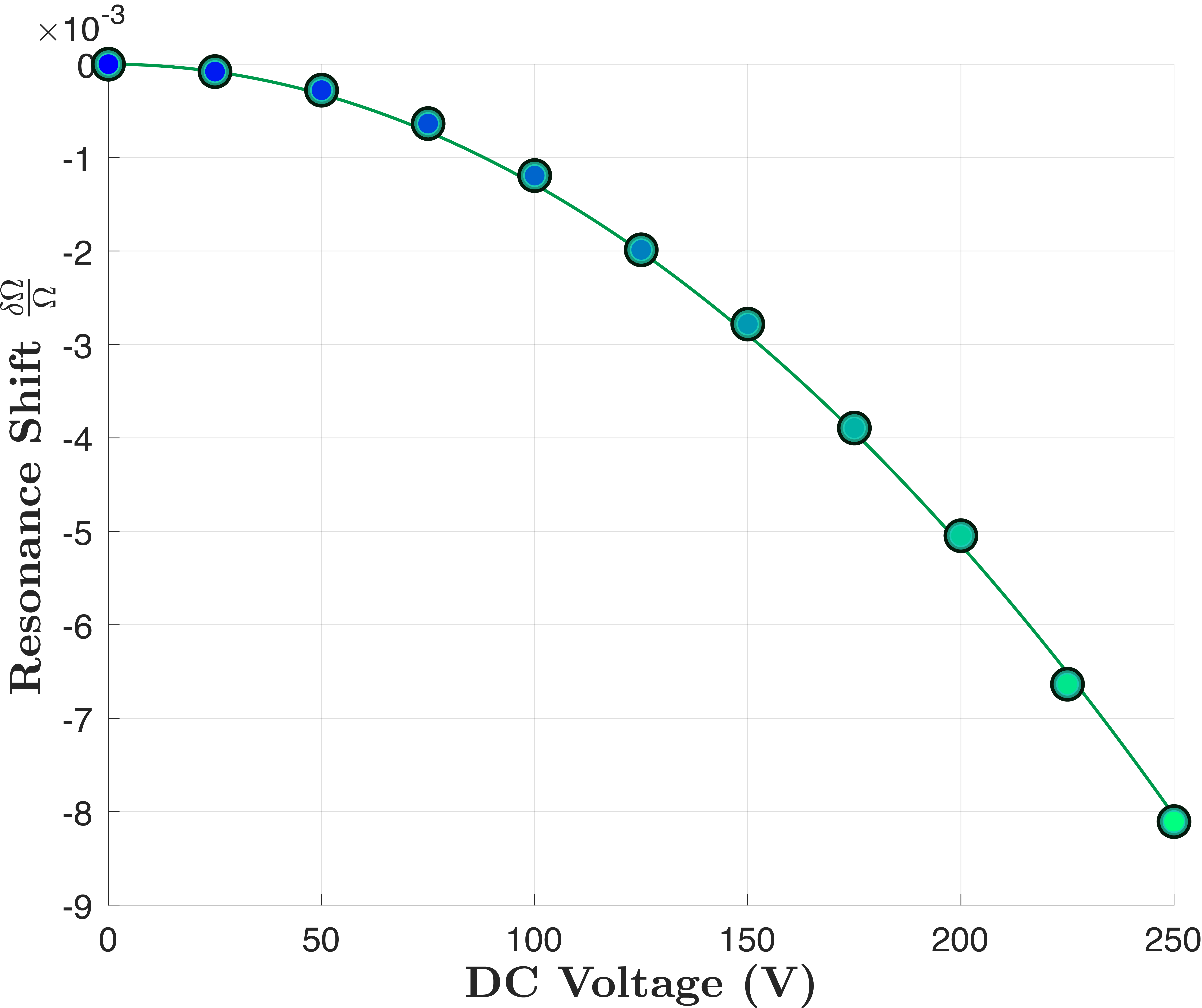}
			\caption{The gold membrane frequency shift of the (0,1) mode due to the electrostatic force, which allowed the calculation of the approximate distance of the electrode to membrane, to be 195 $\mu$m using Eq.~\ref{eq:dist}. 
				\label{fig:mshift}}
		\end{center}
	\end{figure}
	
Outside the microwave cavity there was an aluminium bracket that housed both of the copper electrodes. An additional adapter (not shown) was created to set the distance between the electrodes and the membrane. The copper electrodes were epoxied to the aluminum bracket using Stycast 2850, which is electrically insulating and thermally conducting. We used a 1 mm thick aluminium washer to raise the entire electrode adapter and position the electrodes below the plane of the membrane frame. The distance of the electrode to the membrane was calibrated by fitting a quadratic function to the electrostatic frequency shift (outside the thermal Casimir region) of the fundamental mode, as shown in Fig.~\ref{fig:mshift}. The fit has a parabolic shape, and from a Taylor-expansion of the resonance frequency under the influence of an electrostatic force is given by\cite{pate2018electrostatic,shoaib2016frequency}: 
\begin{equation}
\frac{\delta \Omega_{m}}{\Omega_{m}} \approx \frac{1}{2}\frac{k_{\text{elec}}}{k_{\text{mech}}} = - \frac{\epsilon A}{2d^{3} k_{\text{mech}}}V_{DC}^{2}. \label{eq:dist}
\end{equation}


The microwave frequency for the re-entrant cavity with a frustum was approximated as a lumped LC model\cite{fujisawa1958general,barroso2004reentrant} with the following geometric parameters illustrated in Fig.~\ref{fig:paramdraw}: 
\begin{gather*}
r_0=\bar{r_0}-\frac{x}{\tan \alpha}, \\
L =\frac{\mu_0 h}{2\pi}\left (  \ln \frac{er_2}{r_1} - \frac{r_0}{r_1 - r_0} \ln \frac{r_1}{r_0}  \right ), \\
l_m=\frac{1}{3}\frac{ \sqrt{ \{  2(r_1-r_0)^2 + 3( r_2-r_1)(r_1+r_2-2r_0)  \}^2 + h^2( 3r_2-2r_1-r_0 )^2 } }{2r_2-r_1-r_0}, \\
C_0= \frac{\pi \epsilon_0 \bar{r}_0^2}{x}, \\
C_1 = \frac{\pi \epsilon_0 (r_0^2 - \bar{r}_0^2) }{x} + \frac{2\pi \epsilon_0}{\alpha}\left [  r_0 \ln \frac{e l_m \sin \alpha}{x} + \frac{x \cot\alpha}{2} \ln \frac{\sqrt{e} l_m \sin \alpha}{x}  \right ].
\end{gather*}
 \begin{figure}[!htb]
		\begin{center}
			\includegraphics[width=12cm]{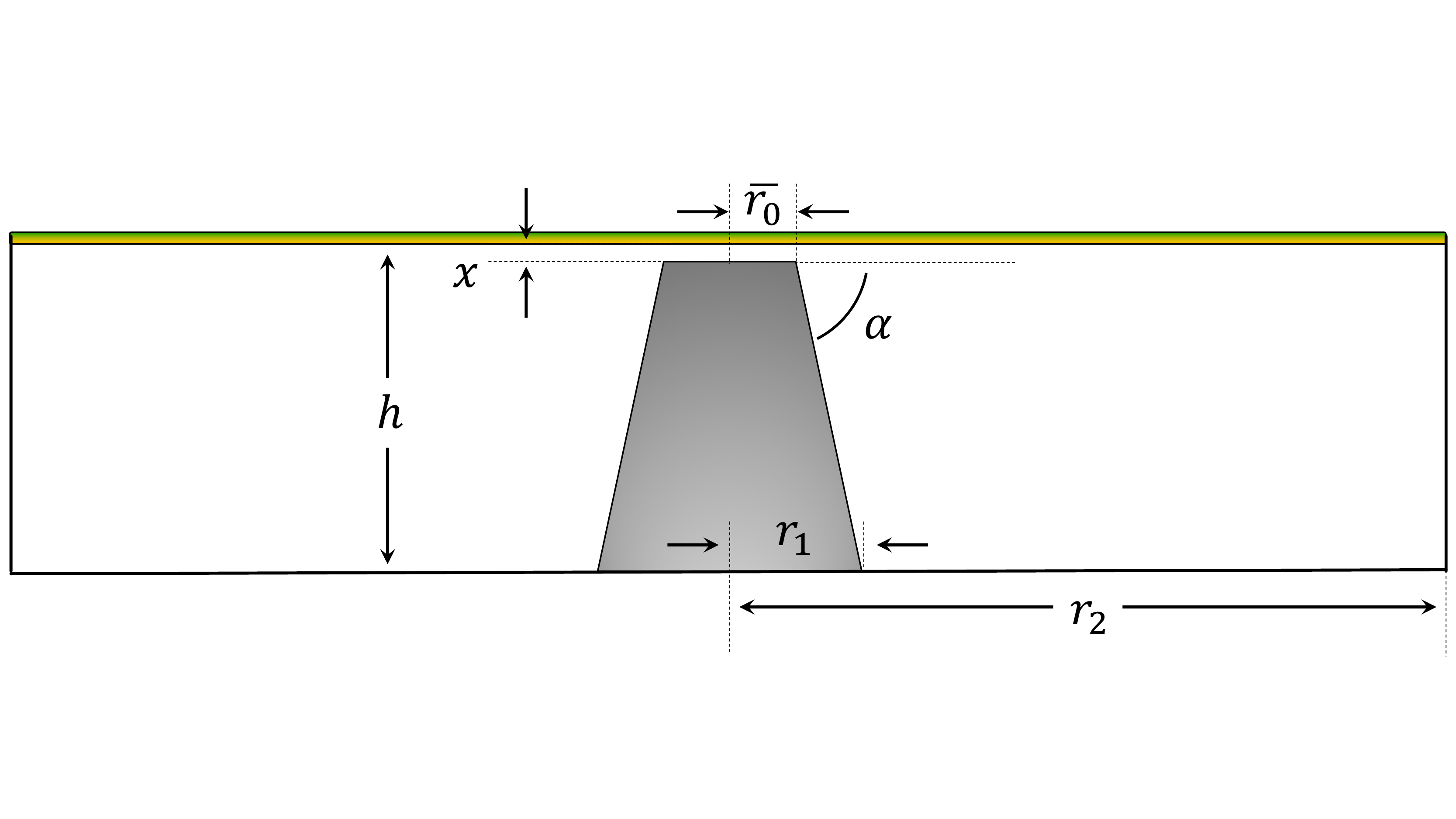}
			\caption{The re-entrant cavity parameters for determining the resonance frequency and its relation to gap size $x$.
				\label{fig:paramdraw}}
		\end{center}
\end{figure}

Based upon these geometric definitions the resonance frequency could be expressed as the familiar equation for an LC oscillator:
\begin{equation}
f=\frac{1}{2\pi \sqrt{L(C_0+C_1)}}.
\end{equation}
In other words, $C_{0}$ is the ``bare capacitance'' while $C_{1}$ can be described as a ``fringing capacitance'' that accounts for the walls of the post. The cavity height $h$ is the sum of the re-entrant post height and the gap size $h=h_{p}+x$. The final geometry of the cavity has the following parameters: $h_{p}=4.25\times 10^{-3}$ m, $\bar{r}_{0}=1.85\times 10^{-4}$ m, $\alpha=87$ degrees, $r_{1}=3.5\times 10^{-4}$ m, and $r_{2}=19.05\times 10^{-3}$ m. 
	


\section{Mechanical Bistability}

In conjunction with the microwave mode hopping presented in Fig.~3 of the main text, the bi-stability of the membrane was recorded in Fig.~\ref{fig:bist} through the observation of the microwave mode hopping. The bias voltage was kept fixed and the frequency of the acoustic drive $V_{ac}$ was swept through the fundamental mechanical resonance of the gold membrane. The maximum frequency separation of the two modes was tracked as $V_{ac}$ passed over the resonance and ``traced'' out the non-linear mechanical response. 
	\begin{figure}[!htb]
		\begin{center}
			\includegraphics[width=12cm]{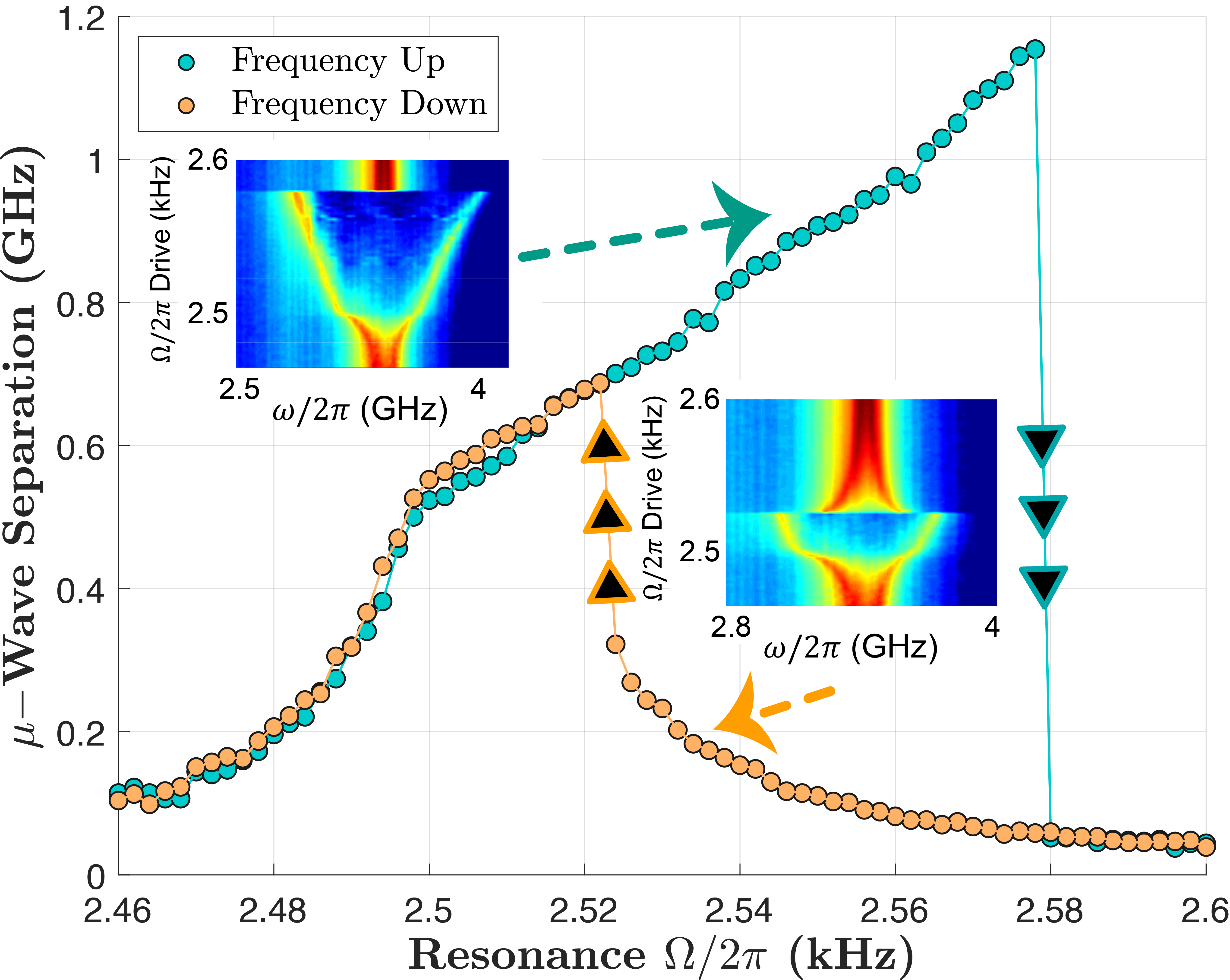}
			\caption{Mechanical bistability of the gold membrane observed using the strength of microwave mode hopping as $V_{ac}$ was swept through the acoustic resonance. Both of the lines were acquired from the maximum separation distance of the two microwave resonances.
				\label{fig:bist}}
		\end{center}
	\end{figure}


\section{Niobium Membranes and Acoustic Response}

Figure~\ref{fig:Nbshift} shows the similar static Casimir spring experiments for the niobium membrane as was shown for the gold membrane in the main text. The cavity experienced a smaller transition between the pinned and free state when compared to the gold membrane, but occurred at a lower frequency of $\omega_{c}/2\pi \approx 2.8$ GHz. The reduced non-linear transition at a lower microwave frequency can be explained by the fact that the niobium membrane was stiffer than the gold membrane, so the balance between the Casimir spring constant and the membrane spring constant occurred at a larger value and hence at a smaller value of the gap spacing.  \\

 \begin{figure}[!htb]
		\begin{center}
			\includegraphics[width=12cm]{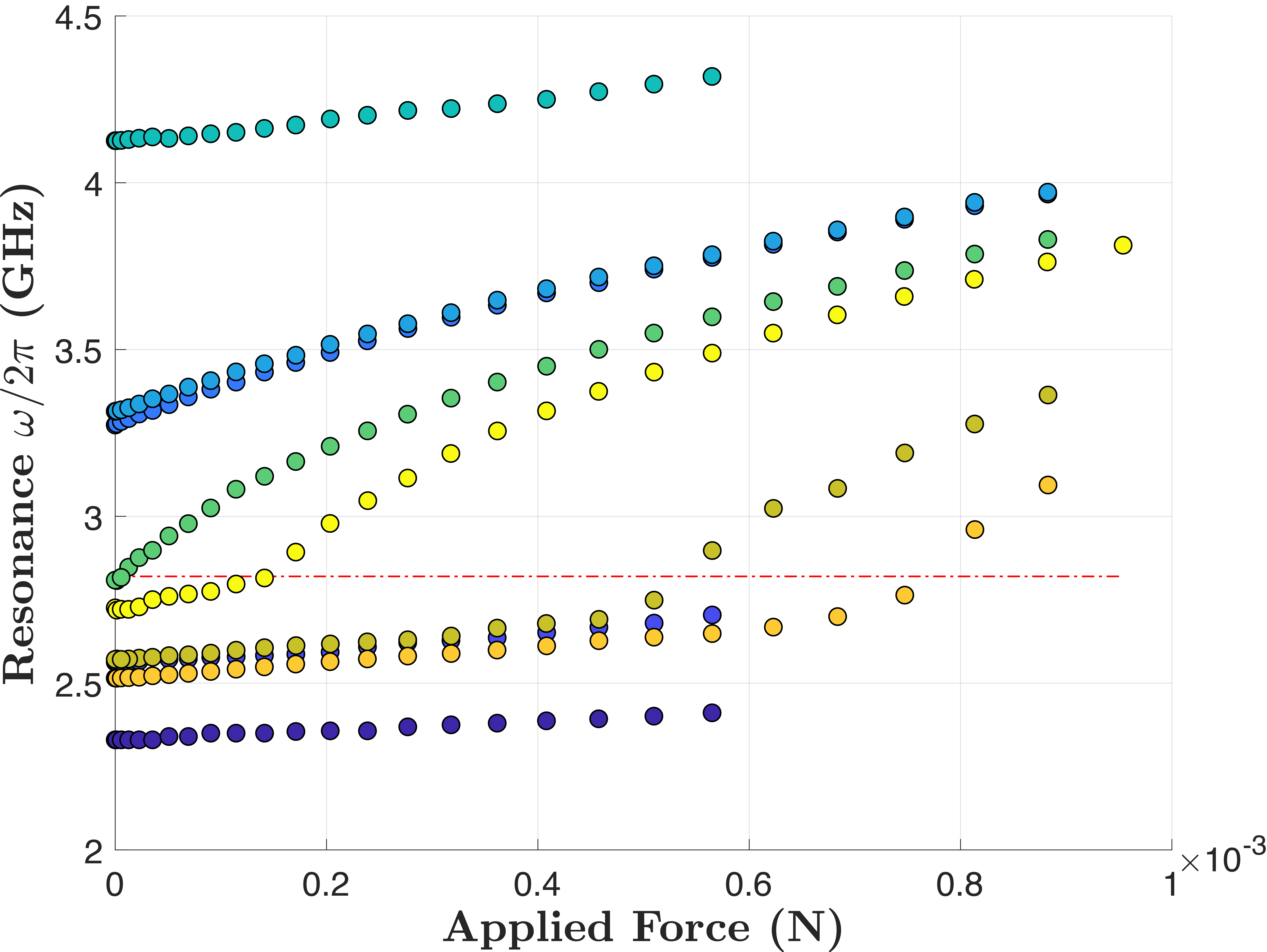}
			\caption{Similar Casimir spring behavior was observed with the niobium membrane cavities as was the case with the gold membrane cavities. Some of the curves do not extend to 250 V because the same voltage power supply was not available as the other curves that extend further. The red line represents a gap size of $x=1.2$ $\mu$m and the point at which the frequency experiences a large shift.
				\label{fig:Nbshift}}
		\end{center} 
	\end{figure}

\begin{figure}[!htb]
	\begin{center}
	\includegraphics[width=12cm]{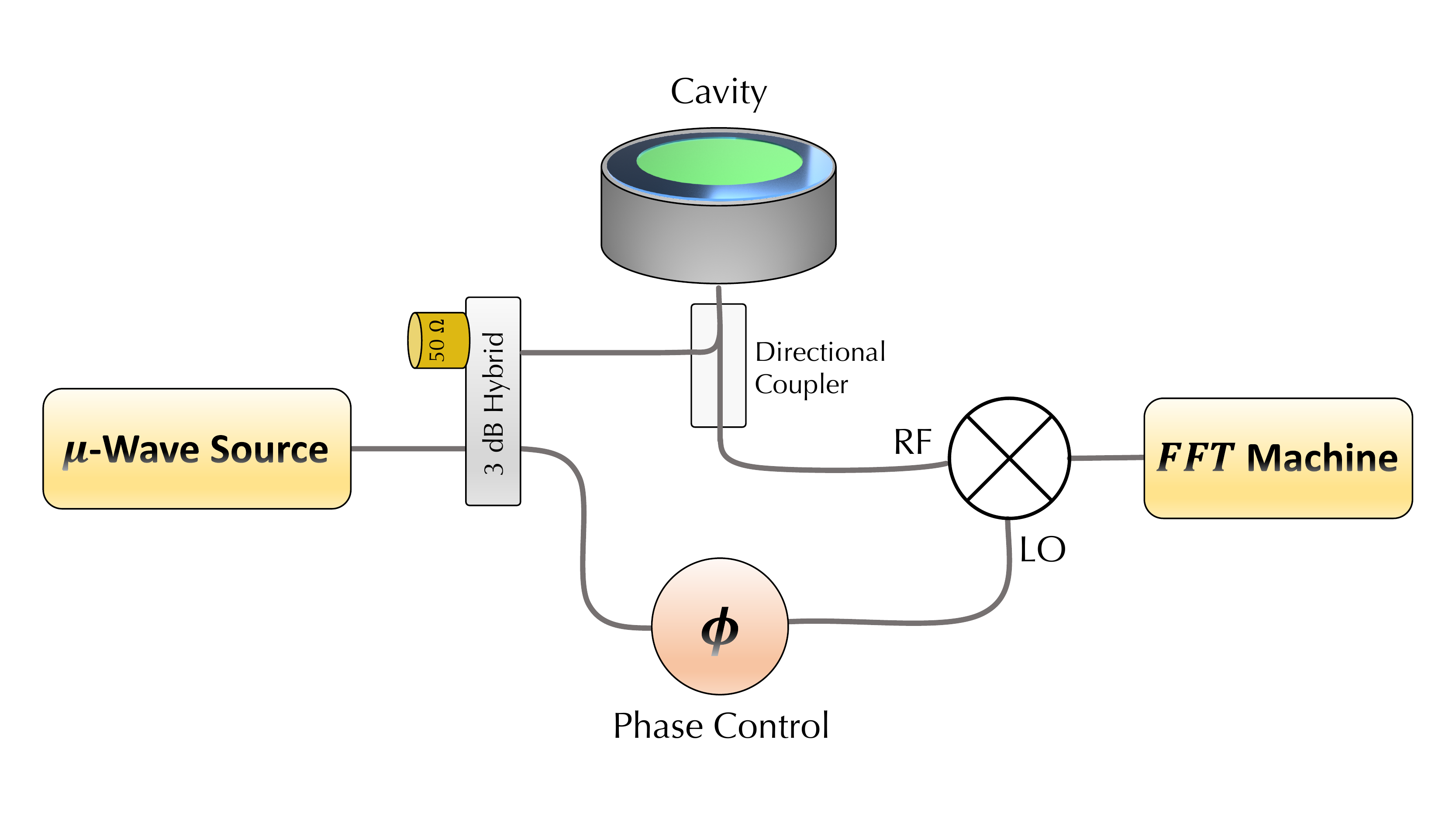}
	\caption{The phase-bridge circuit constructed for measuring the mechanical motion of the membrane. The phase-shift on the cavity is set so at the resonance frequency a maximum voltage response is produced at the output of the mixer.  
	\label{fig:PhaseBridge}}
	\end{center}
\end{figure}

The microwave phase-bridge circuit pictured in Fig.~\ref{fig:PhaseBridge} allowed us to detect the phase modulation of the cavity resonance. The sensitivity of the phase-bridge circuit, $\frac{dV}{df}$, was measured using a known modulation when the phase bridge was tuned to the resonance frequency of the cavity. This was repeated at each point of the tuning that the acoustic quality factor and frequency was measured. The direct measurement of the sensitivity naturally accounted for changes in coupling and electromagnetic $Q$-factor of the microwave cavity, and any other variations in frequency response as the gap spacing was tuned.


	\begin{figure}[!htb]
		\begin{center}
			\includegraphics[width=0.9\textwidth]{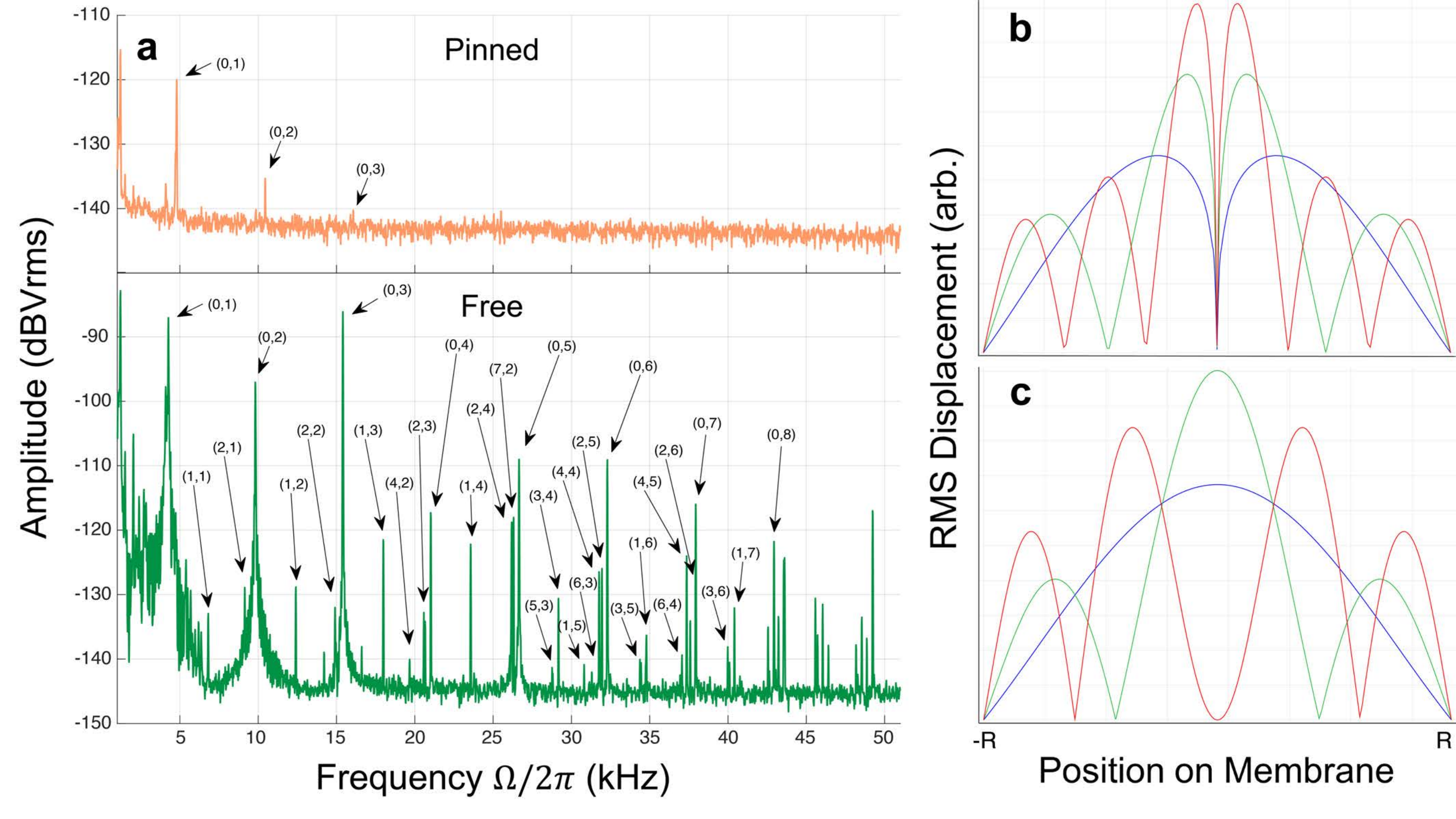}
			\caption{(\textbf{A}) Experimental results of the spectrum at the output of the microwave phase bridge in Fig.~\ref{fig:PhaseBridge} for the niobium membrane for two different re-entrant cavity gap sizes, where the top orange curve is in the pinned state while the bottom green curve is in the free state.  Here, the re-entrant cavity frequencies were measured to be $\omega_{c}/2\pi$=2.51 GHz ($x_0=0.92$ $\mu$m, orange) and $\omega_{c}/2\pi$=3.15 GHz ($x_0=1.50$ $\mu$m, green) respectively. (\textbf{B}) The mode displacement profile as modeled by COMSOL in the strongly-pinned state as a function of radial position for the (0,1), (0,2), and (0,3) modes of the membrane in blue, green, and red, respectively. (\textbf{C}) The same modes as modeled by COMSOL in the free state. Here, the re-entrant post was modeled to exist at the radial position equal to zero, equivalent to the real cavity implemented in experiments. Because the pinned state measurements were made at a displacement node, while in the free state measurements were made at a displacement anti-node, there is a large discrepancy in signal to noise ratio between the two measurements. For example, in the free state some modes are 50 dB above the noise floor where in the pinned state only a few modes were detectible with significantly reduced sensitivity. 
				\label{fig:Nbmechext}}
		\end{center}
	\end{figure}

Fig.~\ref{fig:Nbmechext}A shows the acoustic spectrum as measured by the phase bridge technique in the ``free'' state and ``pinned'' state. There was an increased signal amplitude (up to 50 dB for some modes), frequency shift, and appearance of additional circular drum modes as the membrane transitioned from the pinned into the free state. This was because in the pinned state, the microwave re-entrant cavity measured the acoustic mode near a node, while the free state was near an anti-node, as indicated in Fig.~\ref{fig:Nbmechext} B,C.
	

\begin{table}[!htb]
\caption{\label{tab:table1} Tabulated list of the experimental measured acoustic resonance modes of the niobium membrane in the pinned and free state, identified as acoustic modes using the microwave mode hopping technique. We only show measured modes of up to 27 kHz for the comparison, even though additional modes of up to 50 kHz were found when $\omega_{c}/2\pi \approx 3.15$ GHz. The resonances in parentheses were obtained with the COMSOL simulation for comparison with experimental results.}
\begin{center}
\begin{tabular}{P{6.5cm} P{6.3cm} }
$\omega_{c}/2\pi \approx 2.50$ GHz (Pinned),
Measured Resonances $\Omega_{m}/2\pi$ (kHz) & $\omega_{c}/2\pi \approx 3.15$ GHz (Free), Measured Resonances $\Omega_{m}/2\pi$ (kHz)\\ 
\hline
4.880 (4.851) & 4.141 \\
10.442 (10.501)  & 4.187 \\
16.064 (16.150)  & 4.258 \\
21.661 (21.728) & 4.348 \\
27.291 (27.478) & 4.480 \\
& 9.806 \\
& 9.890 \\
& 12.393 \\
& 12.550 \\
& 13.525 \\
& 14.921 \\
& 14.969 \\
& 15.421 \\
& 17.981 \\
& 18.207 \\
& 20.655 \\
& 21.043 \\
& 23.560 \\
& 23.871 \\
& 26.185 \\
& 26.300 \\
& 26.690 \\
\end{tabular}
\end{center}
\end{table}

The modes of the niobium membrane were identified from the driven response of the membrane, that is, from the observation of microwave mode hopping. This allowed us to distinguish between the real mechanical modes and other forms of narrow band interference. The identified mode frequencies are documented in Table~\ref{tab:table1}. There was a big difference in the acoustic spectrum for the two gap sizes  with microwave resonance frequencies of 2.50 GHz (pinned state) and 3.15 GHz (free state). Only five modes of up to 27 kHz were observed due to the reduced sensitivity for our configuration when in the pinned state. By modeling the membrane with COMSOL, we verified these to be radially-symmetric (0,1) to (0,5) modes. The COMSOL modeling revealed that under the influence of a strong Casimir force, each mode was shifted by an asymptotic amount in the limit of a large Casimir spring (see Fig.~\ref{fig:COMSOL}) $\Omega_{(0,n),\text{cas}}/2\pi \rightarrow \Omega_{(0,n)}/2\pi + \delta_{(0,n)}$. This asymptotic approach is only a manifestation of the fact that the pinning site does not exceed the area of the re-entrant cone. The notation of $\Omega_{0,n}$ signifies that only the radially symmetric ($m=0$) modes in the Casimir region were able to be observed.

A finite-element model in COMSOL was created to see if the Casimir modification to the membrane agrees with the observed mechanical frequency shifts during the transition from the free state to the pinned state for the (0,2) and (0,3) modes as shown in Fig.~\ref{fig:COMSOL}. The simplest approach was to model the Casimir pinning effect as a variable spring at the center point of the membrane. The power law curve of $x^{-4}$ from Fig. 2 of the main text was used to generate a conversion for $k_{eff} \Leftrightarrow x$ in the Casimir region to compare with the COMSOL simulations. The modeled COMSOL data was statically shifted up in effective spring value by identical amounts ($k_{\text{COMSOL,eff}}=k_{\text{COMSOL}}+2350$) for both modes to see the same relative increase in effective spring constant due to the Casimir force. From here, the point at which the membrane frequencies deviated in the Casimir regime for both simulation and real data becomes easier to observe. Good agreement was found with the simulation and measurement for the transitional frequency shifts of the two modes. There appeared to be additional modes near the (0,1) mode fundamental frequency, perhaps global or coupled modes, that made measurement of the frequency shift and determination of $Q_{m}$ more difficult, which was why the higher order modes were chosen. 

		\begin{figure}[!htb]
		\begin{center}
			\includegraphics[width=10cm]{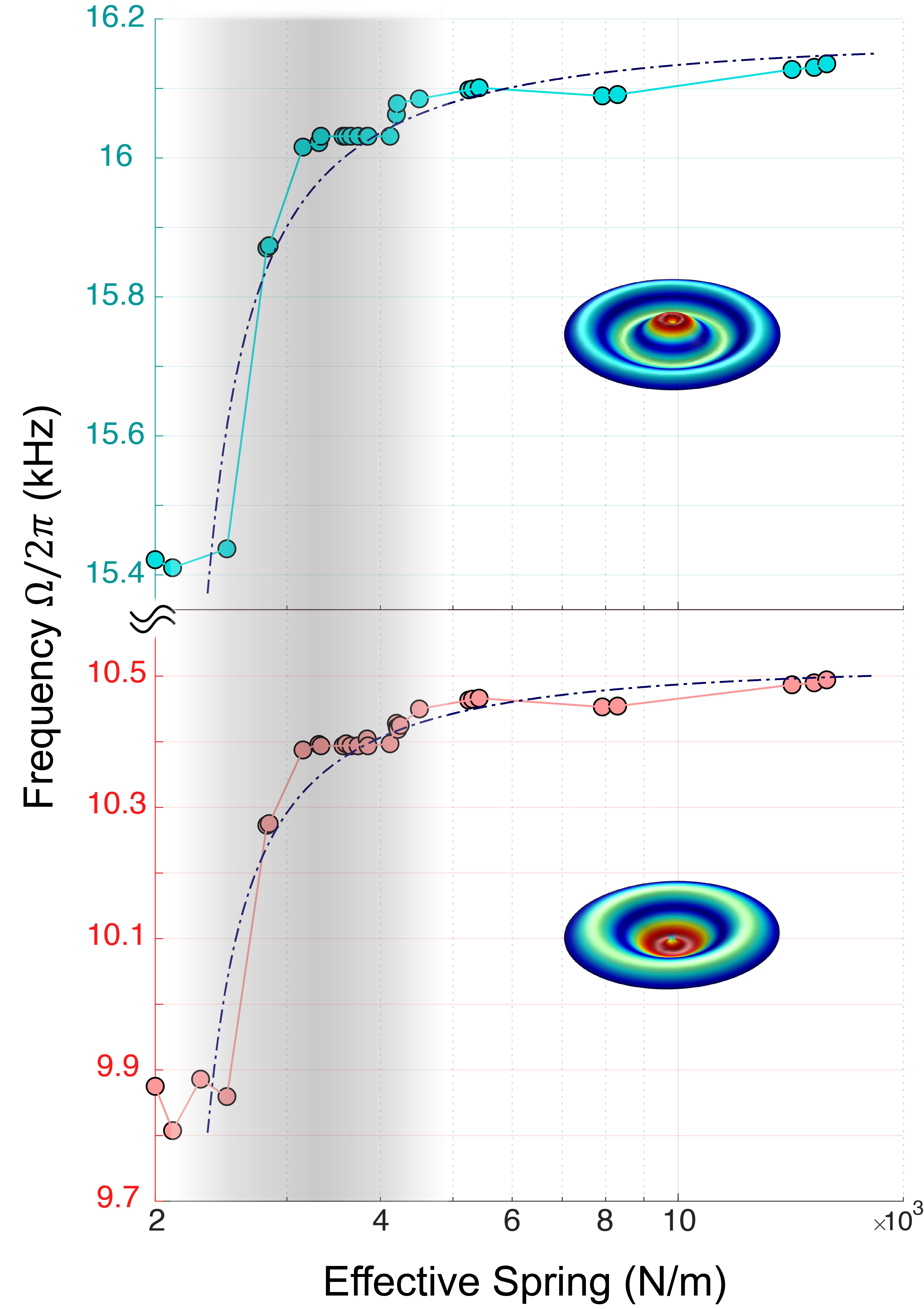}
			\caption{The mechanical resonance frequency shift for the niobium membrane as a function of effective spring. The black line represents the COMSOL model for a variable spring attached to the center of the membrane. The calculated resonance frequency shifts for both (0,3) (upper panel) and (0,2) (lower panel) modes agreed well with increased pinned tension. The shown inset images illustrate the pinned states of the membrane. The shaded region represent regions of non-linear behavior of the membrane. 
				\label{fig:COMSOL}}
		\end{center}
	\end{figure}

\end{widetext}
	
\end{document}